\documentclass[useAMS,usenatbib]{mn2e}

\newcommand{\HI}{$\rm H~{\scriptstyle I}$}

\usepackage[percent]{overpic}
\usepackage{times}
\usepackage{graphicx}
\usepackage{amsmath}
\usepackage{mathabx}
\usepackage{subfigure}
\usepackage{natbib}
\usepackage{txfonts}
\usepackage{journals}

\title[The structure of MW]
{Constraining the Galactic structure parameters with the XSTPS-GAC 
and SDSS  photometric surveys}

\author[B.Q. Chen et al.]
{B.-Q. Chen,$^{1}$\thanks{E-mail:
bchen@pku.edu.cn (BQC); x.liu@pku.edu.cn (XWL).}\thanks{LAMOST Fellow.}
 X.-W. Liu,$^{1,2}$\footnotemark[1]
 H.-B. Yuan,$^3$
 A.C. Robin,$^4$
 Y. Huang,$^1$\footnotemark[2]
 M.-S. Xiang,$^5$\footnotemark[2]
 \newauthor 
C. Wang,$^1$
J.-J. Ren,$^{1,5}$
Z.-J. Tian,$^1$\footnotemark[2]
H.-W. Zhang$^1$
\\
$^{1}$Department of Astronomy, Peking University, Beijing 100871, P.\,R.\,China\\
$^{2}$Kavli Institute for Astronomy and Astrophysics,
Peking University, Beijing 100871, P.\,R.\,China\\
$^{3}$Department of Astronomy, Beijing Normal University, Beijing 100875, P.\,R.\,China\\
$^{4}$Institut Utinam, CNRS UMR6213, OSU THETA, Universit\'e de Bourgogne-Franche-Comt\'e, Observatoire de Besan\c{c}on,  25010 Besan\c{c}on, France  \\
$^{5}$National Astronomy Observatories, Chinese Academy of Sciences, Beijing 100012, P.\,R.\,China
 }

\begin{document}

\date{Accepted ???. Received ???; in original form ???}

\pagerange{\pageref{firstpage}--\pageref{lastpage}} \pubyear{2016}

\maketitle

\label{firstpage}

\begin{abstract}
Photometric data from the Xuyi Schmidt Telescope Photometric Survey of the Galactic Anticentre (XSTPS-GAC) and the Sloan Digital Sky Survey (SDSS) are used to derive the global structure parameters of the smooth components of the Milky Way. The data, which cover nearly 11,000 deg$^2$ sky area and the full range of Galactic latitude, allow us to construct a globally representative Galactic model. The number density distribution of Galactic halo stars is fitted with an oblate spheroid that decays by power law. The best-fit yields an axis ratio and a power law index $\kappa=0.65$ and $p=2.79$, respectively. The $r$-band differential star counts of three dwarf samples are then fitted with a Galactic model. The best-fit model yielded by a Markov Chain Monte Carlo analysis has thin and thick disk scale heights and lengths of $H_{1}=$ 322\,pc and $L_{1}=$2343\,pc, $H_{2}=$794\,pc and $L_{2}=$3638\,pc, a local thick-to-thin disk density ratio of $f_2=$11\,per\,cent, and a local density ratio of the oblate halo to the thin disk of $f_h=$0.16\,per\,cent. The measured star count distribution, which is in good agreement with the above model for most of the sky area, shows a number of statistically significant large scale overdensities, including some of the previously known substructures, such as the Virgo overdensity and the so-called ``north near structure'', and a new feature between 150\degr $< l < $ 240\degr~and $-1$5\degr $< b < $ $-$5\degr, at an estimated distance between 1.0 and 1.5\,kpc. The Galactic North-South asymmetry in the anticentre is even stronger than previously thought.
\end{abstract}


\section{Introduction}
\label{introduction}

One of the fundamental tasks of the Galactic studies is to
estimate the structure parameters
of the major structure components.
\citet{Bahcall1980} fit the observations with two structure components, namely a disk and a halo. 
\citet{Gilmore1983}  introduce a third component, namely a thick disk, 
confirmed in the earliest Besancon Galaxy Model \citet{Creze1983}.
Since then, various methods and observations have been adopted to
estimate parameters of the thin and thick disks and of
the halo of our Galaxy. As the quantity and quality of data available continue to improve over the years,
the model parameters derived have become more precise, numerically. 
Ironically, those numerically more precise results do not converge (see Table~1 of 
\citealt{Chang2011}, Table~2 of \citealt{Lopez2014} and Sect.~5 and 6 of \citealt{Bland2016} for a review).  
The scatters in density law parameters, such as scale lengths,  
scale heights and local densities of these Galactic components, 
as reported in the literature, are rather large.
At least parts of the discrepancies are caused by degeneracy of 
model parameters, which in turn, can be traced back to  the different 
data sets adopted in the analyses. Those differing 
data sets either probe different sky areas 
\citep{Bilir2006a, Du2006, Cabrera2007, Ak2007, Yaz2010, Yaz2015}, 
are of different completeness magnitudes and therefore 
refer to different limiting distances \citep{Karaali2007},
or of consist of stars of different populations of different absolute magnitudes 
\citep{Karaali2004, Bilir2006b, Juric2008, Jia2014}.  
It should be noted that the analysis of \citet{Bovy2012}, using the SEGUE spectroscopic survey, 
has given a new insight on the thin and thick disk structural parameters. This analysis provides estimate 
of their scale height and scale height as a function of metallicity and alpha abundance ratio. However, 
it relies on incomplete data (since it is spectroscopic) with relatively low range of Galactocentric radius as 
for the thin disk is concerned. 

A wider and deeper sample than those employed hitherto may help break the degeneracy 
inherent in a multi-parameter analysis and yield a globally representative Galactic model. 
A single or a few fields are insufficient to break the degeneracy. 
The resulted best-fit parameters, while sufficient for the description of  
the lines of sight observed, may be unrepresentative of the entire Galaxy. For the latter purpose, 
systematic surveys of deep limiting magnitude of all or a wide sky area, such as the 
Two Micron All Sky Survey (2MASS; \citealt{Skrutskie2006}), the Sloan Digital Sky Survey 
(SDSS; \citealt{York2000}), the Panoramic Survey Telescope \& Rapid Response System 
(Pan-Starrs; \citealt{Kaiser2002}) and the GAIA mission \citep{Perryman2001}, 
are always preferred.

Several authors have studied the Galactic structure with 2MASS data at low 
\citep{Lopez2002, Yaz2015} or high latitudes \citep{Cabrera2005, Cabrera2007, Chang2011}. 
\citet{Polido2013} uses the model from \citet{Ortiz1993} and rederive the parameters of this model
based on the 2MASS star counts over the whole sky area. 
However, the survey depth of 2MASS is not quite enough to reach the outer disk and the halo.
The survey depth of SDSS is much deeper than that of the 2MASS. Many authors 
(e.g.  \citealt{Chen2001, Bilir2006a, Bilir2008, Jia2014, Lopez2014})
have previously used the SDSS data to constrain the Galactic parameters. 
Those authors have only made use of a portion of the surveyed fields, 
at intermediate or high Galactic latitudes. 
\citet{Juric2008} obtain Galactic model parameters from the stellar 
number density distribution of 48 million 
stars detected by the SDSS that sample distances from 100\,pc to 20\,kpc and cover 
6500\,deg$^2$ of sky. Their results are amongst those mostly quoted. 
However, in their analysis, they have avoided the Galactic plane. 
So the constraints of their results on the disks, especially the
thin disk, are weak. In their analysis, \citet{Juric2008} have also 
adopted photometric parallaxes assuming that all stars of the same colour
have the same metallicity. Clearly, (disk) stars in different parts of the Galaxy have quite different 
\citep{Ivezic2008, Xiang2015, Huang2015} metallicities, and these variations in metallicities may 
well lead to biases in  the model parameters derived.

 In order to provide a quality input catalog for the LAMOST Spectroscopic Survey of the Galactic Anticentre
(LSS-GAC; \citealt{Liu2014,Liu2015, Yuan2015}), 
a multi-band CCD photometric survey of the Galactic 
Anticentre with the Xuyi 1.04/1.20m Schmidt Telescope  
(XSTPS-GAC; \citealt{Zhang2013,Zhang2014,Liu2014}) 
has been carried out. The XSTPS-GAC photometric catalog contains 
more than 100 million stars in the direction of Galactic anticentre (GAC). It provides an excellent 
data set to study the Galactic disk, its structures and substructures. 
In this paper, we take the effort to constrain the Galactic model 
parameters by combining photometric 
data from the XSTPS-GAC and SDSS surveys.
This is the third paper of a series on the Milky Way study based on the XSTPS-GAC data.  In 
\citet{Chen2014}, we present a three dimensional extinction map in $r$ band. The map has a spatial 
angular resolution, depending on latitude, between 3 and 9\,arcmin and covers the entire XSTPS-GAC 
survey area of over 6,000 deg$^2$ for Galactic longitude 140 $< l <$220\,deg and latitude 40 $< b <$40\,deg. 
In \citet{Chen2015}, we investigate the correlation between the extinction and the \HI~ and CO emission at intermediate 
and high Galactic latitudes ($|b| >$ 10\degr) within the footprint of the XSTPS-GAC, on small and large scales. 
In the current work we are interested in the global, smooth structure of the Galaxy. 

For the Galactic structure, in addition to the global,  smooth major components,
many more (sub-)structures have been discovered, including the inner bars near the Galactic centre
\citep{Alves2000, Hammersley2000, vanLoon2003, Nishiyama2005, Cabrera2008, Robin2012},
flares and warps of the (outer) disk \citep{Lopez2002, Robin2003, Momany2006, Reyle2009, Lopez2014},
and various overdensities in the halo and the outer disk, such as the Sagittarius Stream
\citep{Majewski2003}, the Triangulum-Andromeda \citep{Rocha2004, Majewski2004} and
Virgo \citep{Juric2008} overdensities, the Monoceros ring \citep{Newberg2002,Rocha2003}
and the Anti-Center Stream \citep{Rocha2003,Crane2003, Frinchaboy2004}.
They show the complexity of the Milky Way. 
Recently, \citet{Widrow2012} and \citet{Yanny2013} have 
found evidence for a significant Galactic North-South 
asymmetry in the stellar number density distribution,  exhibiting some 
wavelike perturbations that seem to be intrinsic to the disk.
\citet{Xu2015} show that in the anticentre regions
there is an oscillating asymmetry in the main-sequence star counts 
on either sides of the Galactic plane, in support of the prediction of 
\citet{Ibata2003}. The asymmetry oscillates in the sense that there are more 
stars in the north, then in the south, then back in the north,
and then back in the south  at distances of about 2, 4 -- 6, 8 -- 10 and 12 -- 16\,kpc 
from the Sun, respectively.

The paper is structured as follows. The data are introduced in Section~2.
We describe our model and the analysis method in Section~3. Section~4 
presents the results and discussions. In Section~5 we discuss the large 
scale excess/deficiency  of star counts that reflect the substructures in the halo and disk.  
Finally we give a summary in Section~6.

\section{Data}

\begin{table}
 \centering
  \caption{Data sets.}
  \begin{tabular}{lcccc}
  \hline
  \hline
 & area & field size & $N_{\rm fields}$& $r$ ranges \\
  &  (deg$^2$)    & (deg $\times$ deg)& & (mag) \\
  \hline
XSTPS-GAC & $\sim$3392 & 2.5$\times$2.5 & 574 &12--18 \\
XSTPS-M31/M33 & $\sim$588 & 2.5$\times$2.5 & 108 & 12--18 \\
SDSS & $\sim$6871 & 3.0$\times$3.0 & 1592 &15--21 \\
   \hline
\end{tabular}\\
\end{table}

\begin{figure}
  \centering
  \includegraphics[width=0.48\textwidth]{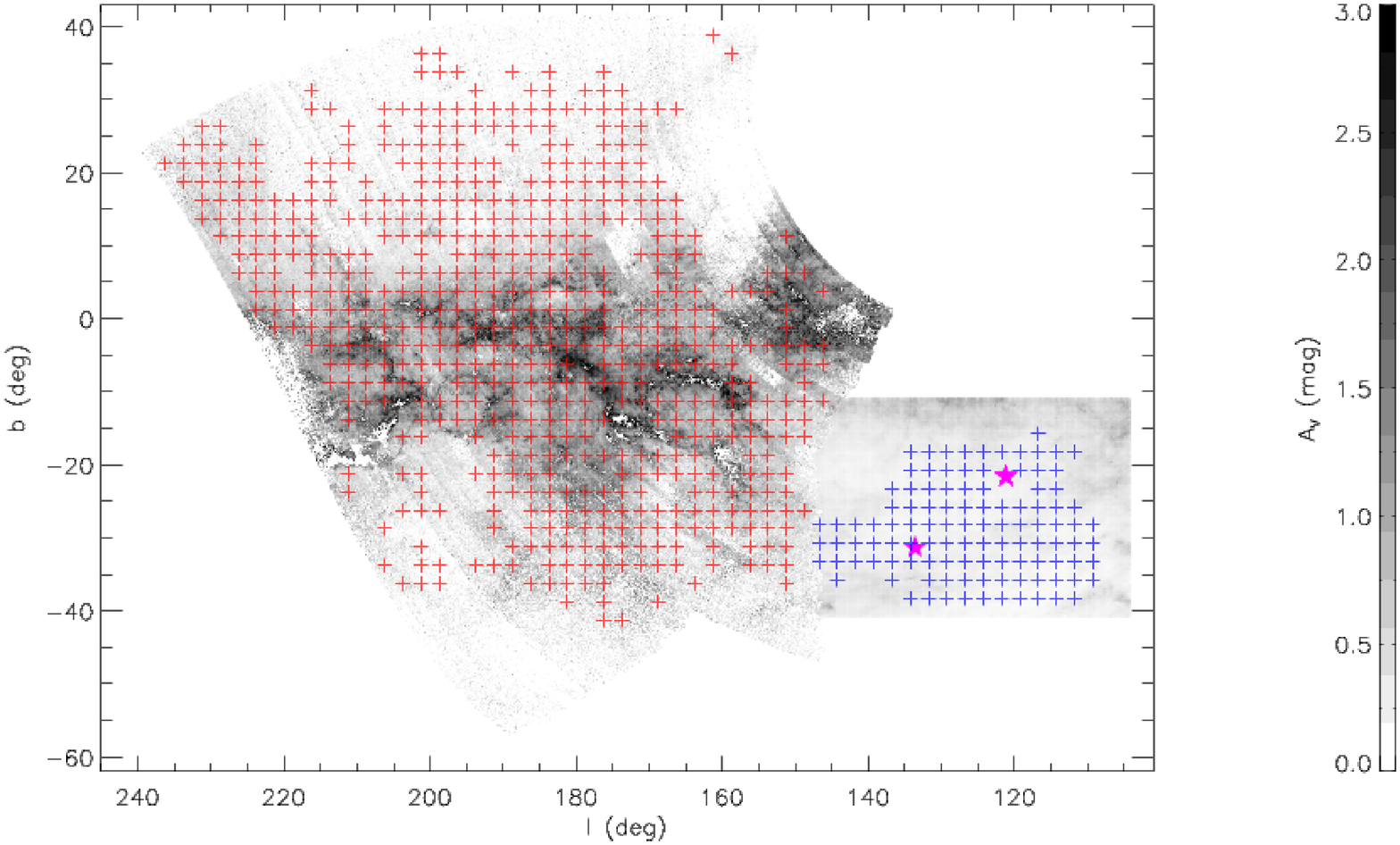}
  \includegraphics[width=0.48\textwidth]{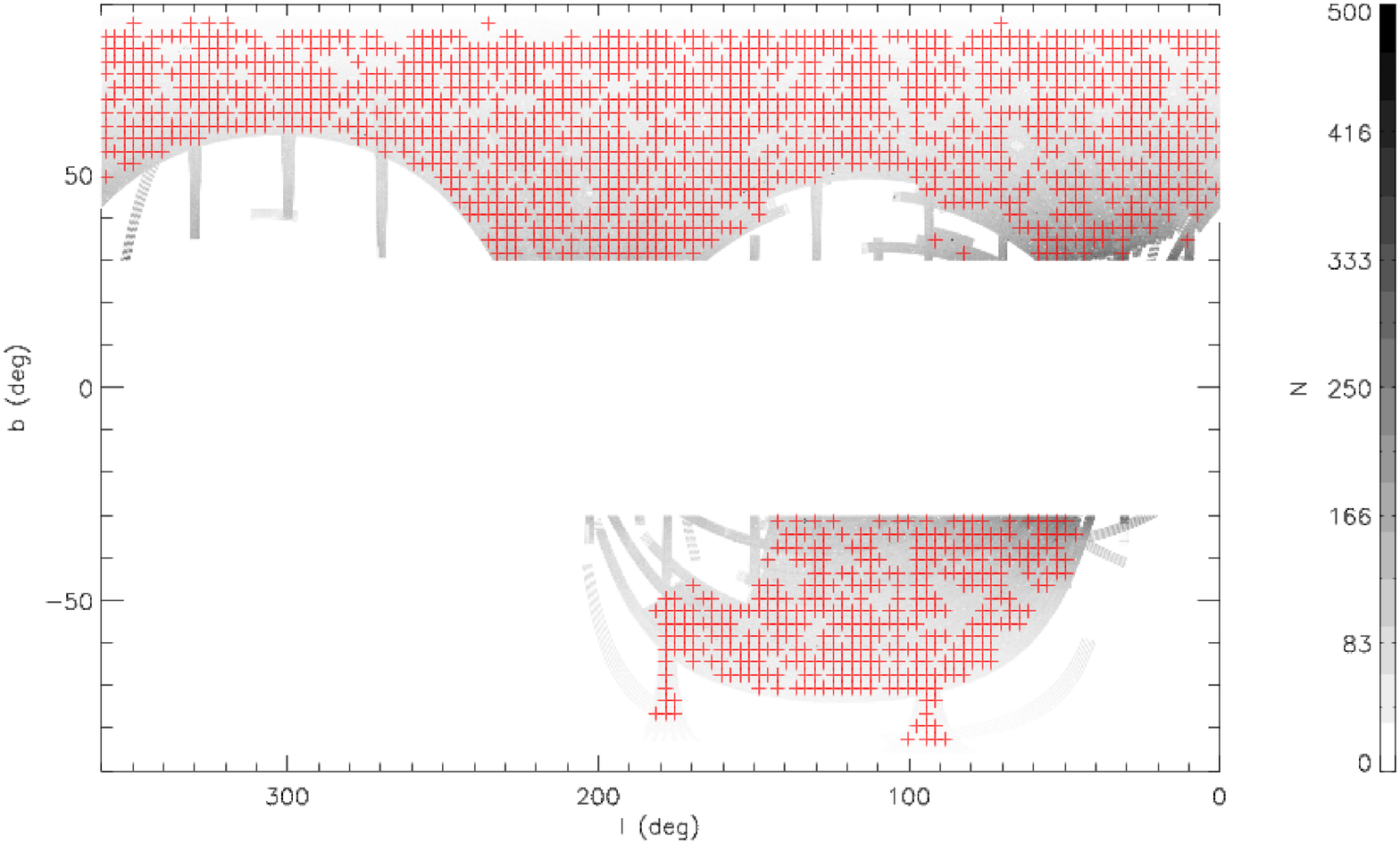}
  \includegraphics[width=0.48\textwidth]{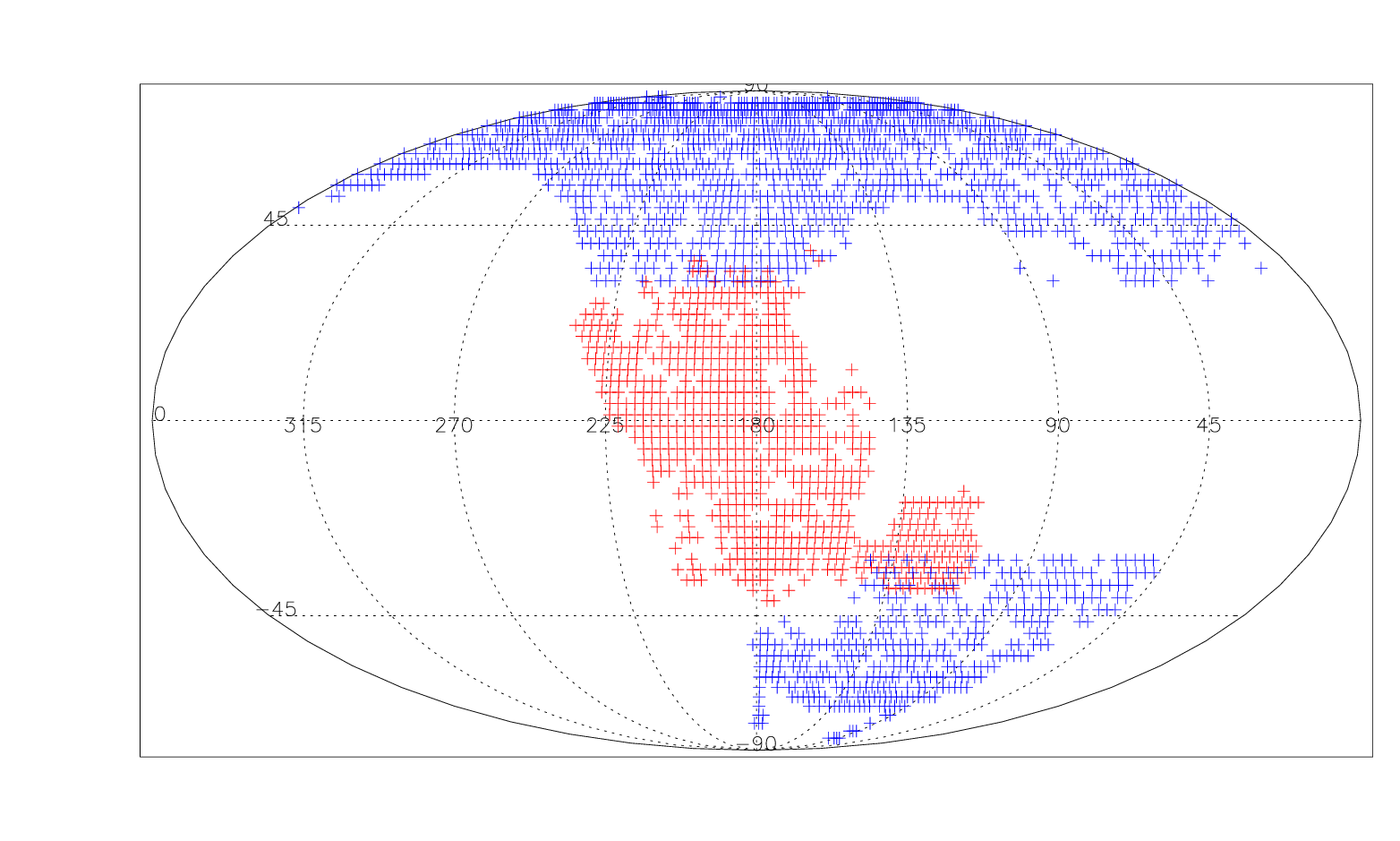}
  \caption{{\it Upper panel}: Extinction map of the GAC and M31/M33 areas within the footprint of 
XSTPS-GAC (\citealt{Chen2015} map for GAC area and \citealt{Schlegel1998} map for M31/M33 area). 
The selected fields for GAC area and M31/M33 area are marked as red and blue pluses,
respectively. The red star symbols mark the central positions of M31 and M33, respectively. {\it Middle panel}: 
SDSS DR12 density map of stars in a magnitude bin of $r$ = 15.5 to 16.5\,mag at a 
resolution of 0.1\degr. The selected fields from SDSS are marked as red pluses.   {\it Bottom panel}: 
Location of the 682 fields selected from the XSTPS-GAC (red) and 1592 fields selected from the SDSS (blue) in 
Galactic coordinates.}
  \label{data}
\end{figure}

\begin{figure*}
  \centering
  \includegraphics[width=0.9\textwidth]{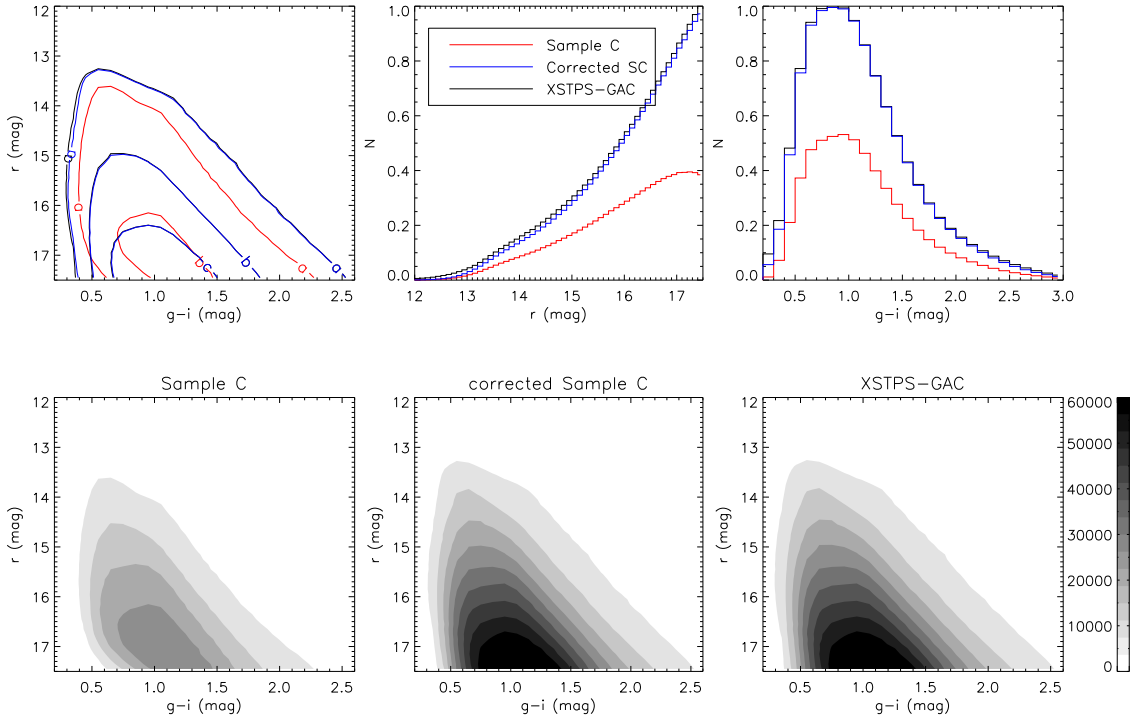}
  \caption{Colour-magnitude distributions of stars in all selected subfields from 
the Sample\,C, the re-weighted Sample\,C (see Equation~(1) 
and related discussion), and the XSTPS-GAC.
The bottom three panels show the grey-scaled number densities distributed in the $g-i$ vs. $r$ space
respectively for the XSTPS-GAC (left), the re-weighted Sample\,C (middle), and the XSTPS-GAC (right).
The upper three panels show the number distribution contours in the $g-i$ vs. $r$ space
(left) as well as number distributions 
respectively in $r$ (middle) and $g-i$ (right) for each sample.
The black contours and histograms show the density of all 
targets in the XSTPS-GAC, the red ones represent the distributions of stars in 
Sample\,C and the blue ones display the 
distributions for the re-weighted Sample\,C. The contours labeled with `a', `b' and `c'
in the left-upper panel represent the contour levels of star number 
of 6\,000,  24\,000 and 48\,000, respectively.
The re-weighted Sample\,C perfectly reproduces the colour-magnitude
sampling provided by the XSTPS-GAC.}
  \label{sel}
\end{figure*}

\subsection{The XSTPS-GAC Data}

The XSTPS-GAC started collecting data in the fall of 2009 and completed in the spring of 2011.
It was carried out in order to provide input catalogue for the LSS-GAC.
The survey was performed in the SDSS $g$, $r$ and $i$ bands using the 
Xuyi 1.04/1.20\,m Schmidt Telescope equipped with a 4k$\times$4k CCD camera, 
operated by the Near Earth Objects Research Group of the Purple Mountain Observatory.
The CCD offers a field of view (FoV) of 1.94\degr $\times$ 1.94\degr, with a pixel scale of 1.705\,arcsec.
In total, the XSTPS-GAC archives approximately 100 million stars down to a
limiting magnitude of about 19 in $r$ band ($\sim$ 10$\sigma$) 
with an astrometric accuracy about 0.1\,arcsec
 and a global photometric accuracy of about 2\% \citep{Liu2014}.
 The total survey area of XSTPS-GAC is close to 7,000\,deg$^2$,
covering an area of $\sim$ 5,400\,deg$^2$ centered on the GAC,
from RA $\sim$ 3 to 9\,h and Dec $\sim~-$10\degr to $+60\degr$, plus
an extension of about 900\,deg$^2$ to the M31/M33 area and the bridging fields 
connecting the two areas.

\subsubsection{GAC area}

In the direction of GAC, the $r$-band extinction exceeds 1\,mag over a significant fraction of the sky 
(see Fig.~\ref{data}). To correct the extinction of stars in high extinguish area 
using extinction maps integrated over lines of sight, 
such as \citet{Schlegel1998}, will introduce over corrections.
It will make stars too bright and blue. We select a subsample, the
so-called ``Sample\,C'' in \citet{Chen2014}, from XSTPS-GAC.
Extinction of all stars in Sample\,C were calculated by the  spectral energy distribution (SED) 
fitting to the multi-band data, including the photometric data from the optical ($g,~r,~i$ from 
XSTPS-GAC) to the near-infrared ($J,~H,~K_S$ from 2MASS and $W1, ~W2$ from the Wide-field 
Infrared Survey Explorer, WISE, \citealt{Wright2010}). 
The extinction of targets in the subsample, Sample\,C, 
is highly reliable, all having minimum SED fitting 
$\chi ^2 _{min} ~<$ 2.0 (see \citealt{Chen2014} for more details). 
We correct the extinction of stars in Sample\,C using the SED fitting extinction
and the extinction law from \citet{Yuan2013}. 
There are more than 13 million stars in Sample\,C. We divide them 
into small subfields of roughly 2.5\degr $\times$ 2.5\degr. The
width ($\Delta l$) and height ($\Delta b$) of each subfield are always exactly 2.5\degr.
Each subfield is not exactly 6.25\,deg$^2$ but varies with Galactic latitude $b$. 
Because of the heavy extinction or poor observational conditions 
(large photometric errors), some subfields have 
obviously small amount of stars, comparing to most normal 
neighboring fields and thus be excluded. 
As a result, 574 subfields, covering about 3392\,deg$^2$, 
are selected. The locations of these subfields are shown in the top panel of 
Fig.~\ref{data}, with the grey-scale background image 
illustrating the 4\,kpc extinction map from \citet{Chen2015}.

For each subfield, Sample\,C does not contain all stars in XSTPS-GAC.
To connect the distribution of targets in Sample\,C
to the underlying distribution of all stars, it is necessary to correct for the effects of the selection 
(often referred to as selection biases). Generally, the selection effects of Sample\,C 
are due to the following two reasons: (1) the procedure by which we cross-match the photometric 
catalogue of the XSTPS-GAC with those of 2MASS and WISE, and (2) the $\chi^2$ cut when we
define the sample with highly reliable extinction estimates.  For the first part, we lose about 15\,per\,cent 
objects, mainly due to the limiting depths of 2MASS and WISE, 
especially at low Galactic latitudes (see Fig.~1 of \citealt{Chen2014}). 
For the second part, we lose more than half of the objects, because of the large photometric errors, 
high extinction effects, or the special targets contamination, 
such as blended or binaries which are not well fitted by the standard SED 
library in \citet{Chen2014}. Our model for the selection function of Sample\,C 
can thus be expressed as the function of the positions ($l$, $b$), colour ($g-i$) 
and magnitude ($r$) of stars, given by,\\
\begin{equation}
  S(l,b,g-i,r) = \frac{N_{\rm SC}(l,b,g-i,r)}{N_{\rm XSTPS}(l,b,g-i,r)},
\end{equation}
where $N_{\rm SC}(l,b,g-i,r)$ and $N_{\rm XSTPS}(l,b,g-i,r)$ are the number of stars 
in the Sample\,C and the XSTPS-GAC, respectively. 
The numbers of objects are evaluated within each subfield with area of $\sim$ 6.25\,deg$^2$, 
each colour ($g-i$) bin ranging from 0 to 3.0\,mag with a bin-size of 0.1\,mag, 
and each $r$-band magnitude bin ranging from 12 to 18.5\,mag with a bin-size of 0.1\,mag. 

The number distributions in colour $(g-i)$ and magnitude $r$ for the stars
in all selected subfields in the Sample\,C, the Sample\,C
re-weighted by the selection effect, as well as the XSTPS-GAC,
are shown as the density grey-scales and 
density contours and histograms in Fig.~\ref{sel}. 
It is clear that our correction of selection effect leads to perfect agreement between the 
complete XSTPS-GAC photometric sample and the re-weighted Sample\,C.

\subsubsection{M31/M33 area}

The dust extinction in the M31 and M33 area is much smaller, 
compared with the GAC area (see the top panel of Fig.~\ref{data}). 
We adopt the extinction map from \citet{Schlegel1998} 
and the extinction law from \citet{Yuan2013} 
to correct the extinction of stars in M31/M33 area. 
Similarly as in the GAC area, all stars in M31/M33 area are divided into small subfields, 
which have width ($\Delta l$) and height ($\Delta b$) always of 2.5\degr. 
We exclude the subfields which have maximum $E(B-V)$ larger 
than 0.15\,mag (i.e. $A_r$ = 0.4\,mag, according to the extinction law from \citealt{Yuan2013}), 
to avoid the relatively large uncertainties caused 
by the high extinction in the highly extinguished regions. The subfields that cover M31 are also 
excluded. As a result, there are 108 subfields in the
M31/M33 area, covering about 588\,deg$^2$. The locations of these subfields 
are also plotted in the top panel of Fig.~\ref{data}, with 
the grey-scale background image illustrating the extinction map from \citet{Schlegel1998}. 
Considering the limiting magnitude of XSTPS-GAC ($r$ $\sim$ 19\,mag), 
we claim that the data in the M31/M33 area from XSTPS-GAC is complete in the magnitude 
range $12 < r_0 < 18$\,mag. 

\subsection{The SDSS Data}

As the survey area of XSTPS-GAC mainly locate around the low Galactic latitudes, 
we also use the photometric data from SDSS, for constraining better the outer disk and 
the halo. We use the photometric data from SDSS data release 12 (DR12, \citealt{Alam2015}). 
The SDSS surveys mainly for high Galactic latitudes, with only a few stripes 
crossing the Galactic plane. It complements one another with the XSTPS-GAC. We cut the 
SDSS data with Galactic latitude $|b| >$ 30\degr, where the influence of the dust 
extinction is small. The dust extinction are corrected using the extinction 
map from \citet{Schlegel1998} and the extinction law from \citet{Yuan2013}.
The SDSS data are divided into subfields with
width ($\Delta l$) and height ($\Delta b$) always of 3\degr. 
To make sure that each subfield is fully sampled by the SDSS survey, we further
divide each subfield into smaller pixels (of size 0.1\degr $\times$ 0.1\degr) and 
exclude the subfield which has no stars detected in at least one of the smaller pixels. 
As a result we have obtained 1592 subfields, covering a sky area of 
about 6871\,deg$^2$. In the middle panel of Fig.~\ref{data}, we show the 
spatial distributions of these subfields,
with grey-scale background image illustrating the number density of the SDSS data.
To remove the contaminations of hot white dwarfs, low-redshift quasars and 
white dwarf/red dwarf unresolved binaries from the SDSS sample, we
reject objects at distances larger than 0.3 mag from the $(r-i)_0$ vs. $(g-r)_0$ 
stellar loci \citep{Juric2008, Chen2014}.
The 95 per\,cent completeness limits of the SDSS images are $u$, $g$, $r$, $i$ and $z$ $=$
 22.0, 22.2, 22.2, 21.3 and 20.5\,mag, respectively \citep{Abazajian2004}. Thus 
the SDSS data is complete in the magnitude range of $15 < r_0 < 21$\,mag. 

A brief summary of the data selection in the current work is given in Table~1. In total, 
there are 2274 subfields, covering nearly 11,000\,deg$^2$,
which is more than a quarter of the whole sky area. 
The positions of all the subfields, from both the XSTPS-GAC and the
SDSS, are plotted in the bottom panel of 
Fig.~\ref{data}. They cover the whole range of Galactic latitudes.
Generally,  the XSTPS-GAC provides nice constraints of the Galactic disk(s), 
especially for the thin disk, while the SDSS provides us a good opportunity to 
refine the structure of Galactic halo, as well as the outer disk.   
 
\section{The Method}

\subsection{The Galactic model}

We adopt a three-components model for the smooth stellar distribution of the
Milky Way. It comprises two exponential disks (the thin disk and the thick disk) 
and a two-axial power-law ellipsoid halo \citep{Bahcall1980, Gilmore1983}. 
Thus the overall stellar density $n(R,Z)$ at a location $(R,Z)$ can be decomposed
by the sum of the thin disk, the thick disk and the halo,
\begin{equation}
  n(R,Z)=D_1(R,Z)+D_2(R,Z)+H(R,Z),
\end{equation}
where $R$ is the Galactocentric distance in the Galactic plane, $Z$ is the
distance from the Galactic mid-plane.
$D_1$ and $D_2$ are stellar densities of the thin disk and the thick disk,
\begin{equation}
  D_i(R,Z)=f_{i}\,n_{0}\exp\left[-\,{(R-R_\odot)\over L_{i}}-\,{(|Z|-Z_\odot)\over H_{i}}\right],
\end{equation}
where the suffix $i=1$ and $2$ stands for the thin disk and thick disk, respectively. 
$R_\odot$ is the radial distance of the Sun to the Galactic centre on the plane, 
$Z_\odot$ is the vertical distance of the Sun from the plane, 
$n_0$ is the local stellar number density of the thin disk at ($R_\odot$, $Z_\odot$),
$f_i$ is the density ratio to the thin disk ($f_1$=1),
$L_{i}$ is the scale-length and $H_{i}$ is the scale-height. 
We adopt $R_\odot=8$\,kpc  \citep{Reid1993} and
$Z_\odot = 25$\,pc \citep{Juric2008} in the current work.
$H$ is the stellar density of the halo,
\begin{equation}
 H(R,Z)=f_{h}\,n_{0}\left[R^2+(Z/\kappa)^2\over R_\odot^2+(Z_\odot/\kappa)^2\right]^{-p/2},
\end{equation}
where $\kappa$ is the axis ratio, $p$ is the power index and $f_h$ is the halo normalization 
relative to the thin disk.

\subsection{Halo fit}

\begin{table}
 \centering
  \caption{The parameter space and results of the halo fit}
  \begin{tabular}{lcccc}
  \hline
  \hline
Parameters & Range & Grid size & Best value  & Uncertainty \\
  \hline
$\kappa$  & 0.1--1.0 & 0.01  & 0.65 & 0.05 \\
$p$  & 2.3--3.3 & 0.01 & 2.79 & 0.17 \\
   \hline
\end{tabular}\\
\end{table}

We fit the component of the halo first.
The metallicity distribution of the halo stars can be described as a single 
Gaussian component, with a median halo metallicity of $\mu_{\rm H}$=$-$1.46\,dex and spatially
invariant of $\sigma_{\rm H}$=0.30\,dex \citep{Ivezic2008}. 
We assume the metallicity of all halo stars as  [Fe/H]$=-1.46$\,dex and adopt the
photometric parallax relation from \citet{Ivezic2008},
\begin{equation}
\begin{split}
M_{r}  = &  4.50-1.11{\rm [Fe/H]}-0.18{\rm [Fe/H]}^2 \\
             &  -5.06+14.32(g-i)_0-12.97(g-i)_0^2 \\
             & + 6.127(g-i)_0^3 - 1.267(g-i)_0^4+0.0967 (g-i)_0 ^5.
\end{split}
\end{equation}
The distances of the halo stars can thus be calculated from the standard relation, 
\begin{equation}
  d=10^{0.2(r_0-M_r)+1}.
\end{equation}

Star in a blue colour bin $0.5 \le g-i < 0.6$ are selected. They do not suffer 
from the giant star contamination and probe larger distances to constrain the halo.
We calculate their distance using Equations~(5) and (6). The distances of the disk stars will be
underestimated because they are more metal-rich. To exclude the contamination of the disk stars, 
we use stars with absolute distance to the Galactic plane 
$|Z| >$ 4\,kpc. For each subfield, we divide all halo stars
into suitable numbers of logarithmic distance bins and then count the 
number for each bin. This number can be modelled as,
\begin{equation}
N_{\rm H}(d)=H(d)\Delta V(d),
\end{equation}
where $H(d)$ is the halo stellar density given by Equation~(4) and  
$\Delta V(d)$ is the volume, given by,
\begin{equation}
\Delta V(d) =\frac{\omega}{3}(\frac{\pi}{180})^2(d^3_2-d^3_1),  
\end{equation}
where $\omega$ denotes the  area of the field (unit in deg$^2$), $d_1$ and $d_2$ are
the lower distance limit and upper distance limit of the bin, respectively. 

We fit the halo model parameters $p$ and $\kappa$ to the data.
As we explicitly exclude the disk, we cannot fit for 
the halo-to-thin disk normalization $f_h$.
A maximum likelihood technique is adopted to explore the best 
values of those halo model parameters.
In Table~2, we list the searching parameter space and the grid size. 
For each set of parameters, a reduced likelihood is computed between the simulated 
data (star counts in bins of distances) and the observations,
given by \citet{Bienayme1987} and \citet{Robin2014},
\begin{equation}
  Lr=\sum_{i=1}^{N} q_i \times (1-R_i+{\rm ln}(R_i)),
\end{equation}
where $Lr$ is the reduced likelihood for a binomial statistics, 
$i$ is the index of each distance bin, $f_i$ and $q_i$ are the 
number of stars in the $i$th bin for the model 
and the data, respectively and $R_i=f_i/q_i$. 
The uncertainties of the halo parameters are estimated similarly as those in \citet{Chang2011}.
We calculate the likelihood for 1000 times using the observed data  and the 
simulations of the best-fit model adding with the Poisson noises. 
The resulted likelihood range defines the confidence level and thus the uncertainties.  

\subsection{Disk fit}

The metallicity distribution of the disk is more complicated than that of the halo. 
Thus we fit the disk model parameters through a different way.
We compare the $r$-band differential star counts in different colour bins and compare 
them to the simulations to search for the best disk model parameters
($n_0, ~L_1,~H_1,~f_2,~L_2$ and $H_2$), as well 
as the halo-to-thin disk normalization $f_h$.

Towards a subfield of galactic coordinates ($l,~b$) 
and solid angle $\omega$, the $r$-band differential star counts $N_{\rm sim} (r^k_0)$ 
($k$ is the index of each magnitude bin) in a given colour bin $(g-i)^j_{0}$ 
($j$ is the index of each colour bin) can be simulated as follows:
\begin{enumerate}
\item The line of sight is divided into many small distance bins. For a given distance 
bin with centre distance of $d_i$ ($i$ is the index of each distance bin),
the $r$-band apparent magnitude of a star is given by 
\begin{equation}
  r_0(d_i) = M_r ((g-i)^j_0, {\rm [Fe/H]}|l,b,d_i)+\mu,
\end{equation}
where $\mu$ is the distance modulus [$\mu = 5{\rm log}_{10}(d_i)-5$] and $M_r$
is the $r$-band absolute magnitude of the star given by Equation (5). The metallicities 
of halo stars are again assumed to be $-$1.46\,dex and those of disk stars are 
given as a function of positions, which is fitted using the metallicities of 
main sequence turn off stars from LSS-GAC \citep{Xiang2015},
\begin{equation}
 {\rm  [Fe/H]} = -0.61+0.51 \cdot {\rm exp}{(-|Z|/1.57)}.
\end{equation}

\item The number of stars in each distance bin can be calculated by,
\begin{equation}
  N(d_i)=n(R,Z|l,b,d_i)V(d_i), 
\end{equation}
where $V(d_i)$ is the volume given by Equation~(8) and $n(R,Z|l,b,d_i)$
is the stellar number density given by Equation~(2, 3 and 4). The halo model parameters,
$\kappa$ and $p$, resulted from the halo fit are adopted and settled to be not changeable here.

\item Combining all distance bins, 
we can obtain the modeled $r$-band star counts $N(r^k_0)$, by
\begin{equation}
  N(r^k_0) = \Sigma N(d_i) ~  {\rm where} ~ r^k_0 - \frac{\rm rbin}{2} < r_0(d_i) < r^k_0 + \frac{\rm rbin}{2},
\end{equation}
where rbin is the bin size of $r$-band magnitude (we adopt rbin=1\,mag in the current work).
$N(r^k_0)$ is the underlying star counts. When comparing to the observations, we 
need to apply the selection function, by
\begin{equation}
  N_{\rm sim}(r^k_0) = N(r^k_0)S(l,b,g-i,r) C,
\end{equation}
where $S(l,b,g-i,r)$ is the selection function, calculated by Equation~(1) for 
XSTPS-GAC subfields in GAC area and equals to one for 
XSTPS-GAC subfields in M31/M33 area and all the SDSS subfields. Besides,
 \begin{equation}
   C =
\begin{cases}
1       & {\rm for~} d_{\rm min} < d_i < d_{\rm max}; \\
0		&  {\rm otherwise};  \\
\end{cases}
\end{equation}
\begin{eqnarray}
  d_{\rm min} &=&  10^{0.2(r_{\rm min}-A_r(d_i)-M_r((g-i)_0,{\rm [Fe/H]}))+1},\\
  d_{\rm max}&=&10^{0.2(r_{\rm max}-A_r(d_i)-M_r((g-i)_0,{\rm [Fe/H]}))+1},
\end{eqnarray}
where $r_{\rm min}$ and $r_{\rm max}$ are the 
magnitude limits of each subfield.
We adopt $r_{\rm min} = 12$ and  $r_{\rm max} = 18$ for all XSTPS-GAC subfields,
and $r_{\rm min} = 15$ and  $r_{\rm max} = 21$ for all SDSS subfields. 
$A_r(d_i)$ is the extinction in $r$-band at distance of $d_i$. We adopt the 3D extinction map from
\citet{Chen2014} for XSTPS-GAC subfields in GAC area and 
2D extinction map from \citet{Schlegel1998} for
XSTPS-GAC subfields in M31/M33 area and all SDSS subfields. As the size of each subfield is quite
large ($\sim$ 4\,deg$^2$), the extinction $A_r(d)$ varies within a subfield. We thus adopt the maximum 
values to make sure that our data are complete.
\end{enumerate}

The photometric parallax relation of Equation~(5) is only valid for the single 
stars. A large fraction of stars in the Milky Way are
actually binaries (e.g. \citealt{Yuan2015b}). In the current work we adopt the binary fraction 
resulted from \citet{Yuan2015b} and assume that 40\,per\,cent of
the stars are binaries. The absolute magnitudes $M_r$ of the binaries are calculated as the same way
as in \citet{Yuan2015b}. 

We also consider the effects of photometric errors, 
the dispersion of disk star metallicities and the errors due to
the photometric parallax relation of \citet{Ivezic2008}. The $r$-band photometric errors
of most stars in the XSTPS-GAC and the SDSS are  smaller than 0.05\,mag 
(\citealt{Chen2014} for the XSTPS-GAC and \citealt{Sesar2006} for the SDSS).
When we fit the metallicities of disk stars as a function of positions [Equation~(11)], we 
find a dispersion of the residuals of about 0.05\,dex. 
According to Equation~(5), this dispersion 
would introduce an offset of about 0.05\,mag for 
the absolute magnitude when [Fe/H] = $-$0.2\,dex. 
As a result, the effect of the photometric errors and the
disk stars metallicities dispersions would 
introduce a distance errors of smaller than 5\,per\,cent.
Combining with the systematic error of the photometric parallax 
relation, which is claimed to be smaller than 10\,per\,cent \citep{Ivezic2008},
we assume a total error of distance of 15\,per\,cent. This distance error is added 
when we model the $r$-band
magnitude of stars in a given distance bin [Equation~(10)].

We select three different colour bins for the disk fit. 
Two of them correspond to G-type stars with 
$0.5 \le (g-i)_0 < 0.6$\,mag and $0.6 \le  (g-i)_0 < 0.7$\,mag,
and the other one corresponds to late K-type stars with $1.5 \le (g-i)_0 < 1.6$\,mag. 
The giant and sub-giant contaminations for the first two G-type star bins are very small. 
For the late K-type stars, we exclude stars with $r$-band magnitude $r_0 < 15$\,mag 
to avoid the giant contaminations. For each colour bin, we count the differential
$r$-band star counts with a binsize of $\Delta r=1$\,mag and then 
compare them to the simulations to search for the 
best disk model parameters, i.e $n_0,~L_1,~H_1,~f_2,~L_2,$ and $H_2$ and
the halo-to-thin disk normalization $f_h$. 
Similarly as in \citet{Robin2014}, 
an ABC-MCMC algorithm is implemented using the reduced likelihood calculated by Equation~(9) 
in the Metropolis-Hastings algorithm acceptance ratio 
\citep{Metropolis1953, Hastings1970}. 
We note that the 68\,per\,cent probability intervals of the 
marginalised probability distribution functions (PDFs) of each parameter, given
by the accepted values after post-burn period in the MCMC chain are only the fitting uncertainties 
which do not include systematic uncertainties.
A detailed analysis of errors of the scale parameters will be given in Sect.~4.2.

The stellar flare is becoming significant at $R \ge$15\,kpc \citep{Lopez2014} 
while the limiting magnitude we adopt for XSTPS-GAC is $r=18$\,mag, which corresponds
to $R ~\sim$ 13\,kpc for early G-type dwarfs.  On the other hand, the disk warp is 
a second order effect on the star counts and the XSTPS-GAC centre around the 
GAC, with $l$ around 180\degr. The effect of the disk warp is thus
negligible \citep{Lopez2002}.  So in the current work we ignore the influences of the disk warps and flares. 
In order to minimise the effects coming from other irregular structures (overdensities)
of the Galactic disk and halo (e.g., Virgo overdensity, etc. ), we iterate our fitting 
procedure to automatically and gradually 
remove pixels contaminated by unidentified irregular structures, similarly as in \citet{Juric2008}. 
The model is initially fitted using all the data points.
The resulted best-fit model is then used to define the outlying data, 
which have ratios of residuals (data minus the best-fit model) to the best-fit model
higher than a given value, i.e. $(N_{\rm obs}-N_{\rm mod})/N_{\rm mod} > a_1$. 
The model is then refitted with the outliers excluded. 
The newly derived best-fit model is again compared to all the data points. 
New outliers with $(N_{\rm obs}-N_{\rm mod})/N_{\rm mod} > a_2$
are excluded for the next fit. We repeat this procedure with a sequence of values
$a_i$ = 0.5, 0.4, and 0.3.  The iteration, which gradually reject about 1, 5 and 15\,per\,cent of
the irregular data points of smaller and smaller significance, will make our model-fitting algorithm to 
converge toward a robust solution which describes the smooth background best. 

\section{The Results and Discussion}

\begin{figure}
  \centering
  \includegraphics[width=0.48\textwidth]{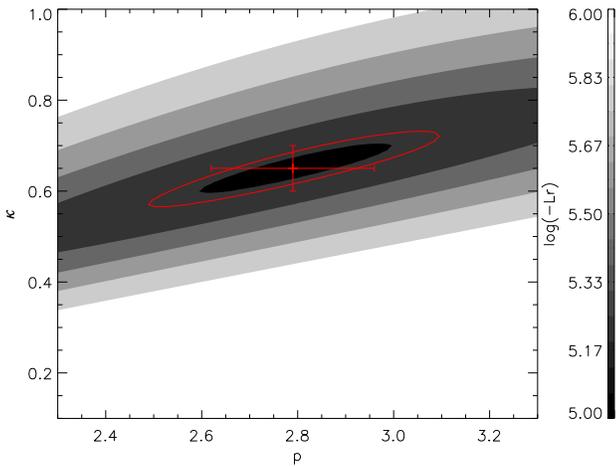}
  \caption{Reduced likelihood surface of the halo parameters $p$ and $\kappa$ 
  space (see Table~2). The best-fitted values and uncertainties are marked 
  as a red plus with error bars. 
  The red contour ellipse shows the likelihood ranges used for estimating 
  the uncertainties.}
  \label{halog}
\end{figure}

\begin{table*}
 \centering
  \caption{The best-fit values of the disk fit}
  \begin{tabular}{lcccccccc}
  \hline
  \hline
  Bin & $n_1$ & $L_1$ & $H_1$ & $f_2$ & $L_2$ & $H_2$ & $f_H$ & $Lr$  \\
         & $10^{-3}$stars\,$pc^{-3}$ & pc & pc & per\,cent & pc & pc & per\,cent &  \\
\hline
Joint fit\\
\hline
 $0.5 \le (g-i)_0 < 0.6$    &  1.25 & 2343 & 322 & 11 & 3638 & 794 & 0.16  & $-$86769   \\
    $0.6 \le (g-i)_0 < 0.7$         &  1.20 & & & & & & &   \\
       $1.5 \le (g-i)_0 < 1.6$           &  0.54 & & & & &  & & \\ 
        \hline    
    Individual fit\\
    \hline
 $0.5 \le (g-i)_0 < 0.6$ & 1.31 & 1737 & 321  & 14   
                               & 3581& 731  & 0.16   & $-$43699   \\     
 $0.6 \le (g-i)_0< 0.7$ & 1.65  & 2350  & 284  & 7  
                               & 3699  & 798  & 0.12   & $-$35774  \\        
 $1.5 \le (g-i)_0 < 1.6$ & 0.41  & 2780  & 359  & 8 
                               & 2926  & 1014  & 0.50 & $-$4028   \\
 \hline
 stddev & & 429 & 31 & 3 & 360 & 124 & 0.02 &  \\
 \hline 
\end{tabular}
\end{table*}     

\begin{figure*}
  \centering
  \includegraphics[width=0.95\textwidth]{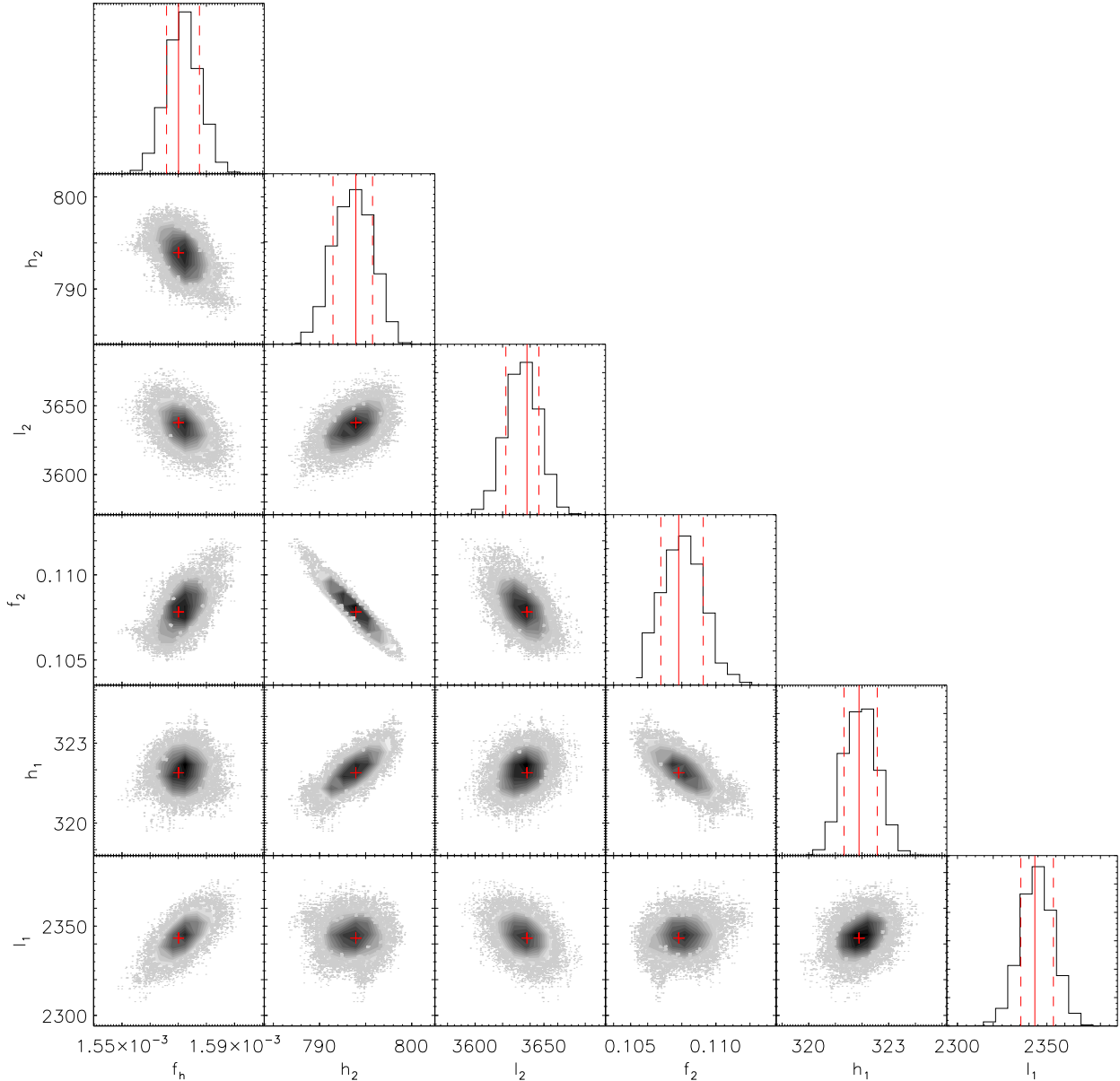}
  \caption{
  Two-dimensional marginalized PDFs for the disk model parameters,
$L_1,~H_1,~f_2,~L_2,$ and $H_2$
and the halo-to-thin disk normalization $f_h$, obtained from the MCMC analysis. 
Histograms on top of each column show the one-dimensional marginalized PDFs 
of each parameter labeled at the bottom of the column. 
Red pluses and lines indicate the best solutions. 
The dash lines give the 16th and 84th percentiles, 
which denotes only the fitting uncertainties.}
  \label{cross2d}
\end{figure*}

\begin{figure*}
  \centering
  \includegraphics[width=0.68\textwidth]{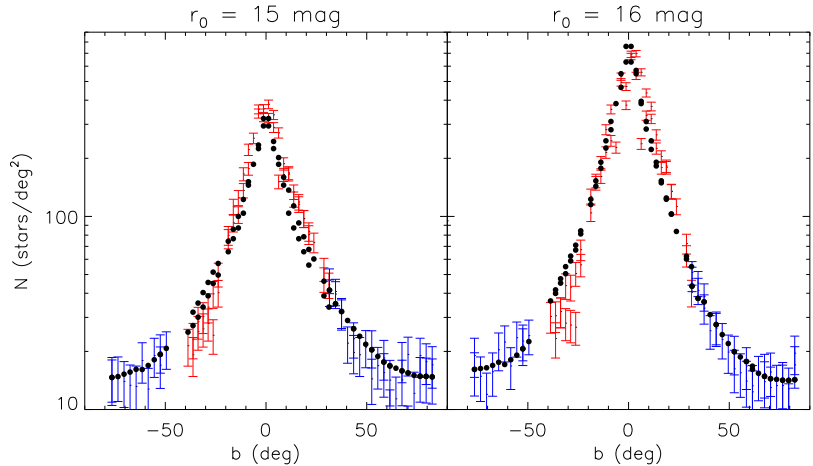}
  \caption{Star count (per deg$^2$) for the colour bin $0.5 \le (g-i)_0 <0.6$\,mag and magnitude bins,
   $r_0$ = 15 (left) and 16\,mag (right), of both the XSTPS-GAC
  (red pluses) and the SDSS (blue pluses) data as a function of the Galactic latitude
for example subfields with Galactic longitude 177\degr  $<l<$ 183\degr.
The black dots are the model predictions.}
  \label{rnmodel}
\end{figure*}

\begin{figure*}
  \centering
  \includegraphics[width=0.7\textwidth]{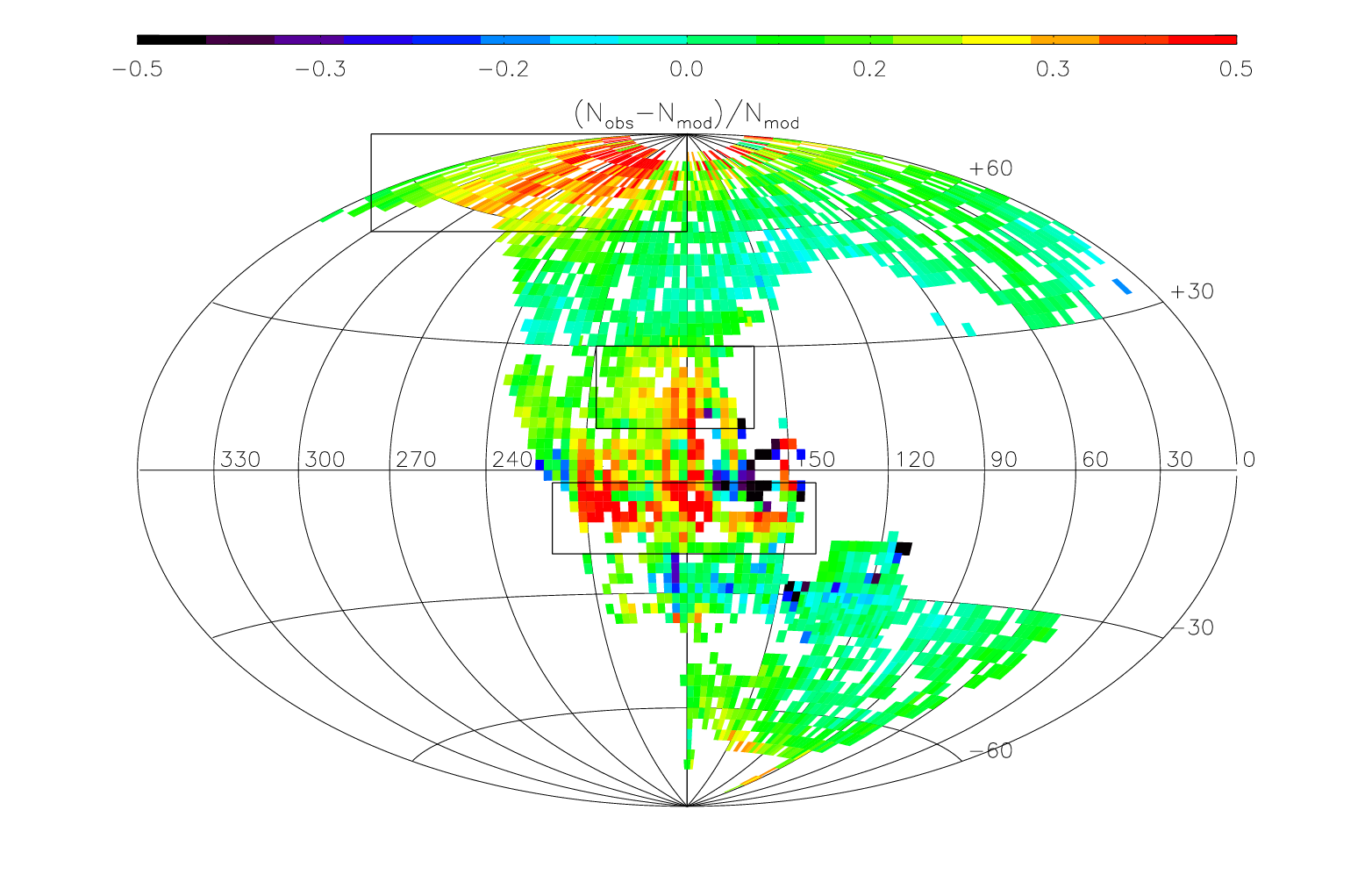}
  \caption{Difference map for the all subfields 
  $(N_{\rm obs}-N_{\rm mod})/N_{\rm mod}$. 
  The populations in excess in the data are most 
  likely irregular structures of the Milky Way.
The squares denote different regions with significant excesses of the
star counts.}
  \label{resd}
\end{figure*}

\subsection{Fitting results}

The best-fit halo model parameters obtained from the halo fit are listed in Table~2 and the
reduced likelihood $L_r$ surface for the fit is shown in Fig.~\ref{halog}.
The $L_r$ contour dramatically changes along $\kappa$, 
but relatively mildly along $p$. The best-fit halo model parameters are 
$\kappa=0.65\pm 0.05$ and $p=2.79\pm 0.17$. They are not 
surprisingly in good agreement with those values from \citet{Juric2008}, 
which have $\kappa=0.64$ and $p=2.8$, since the data used for the halo fit in the current work
are mainly from the SDSS subfields. In addition, the photometric parallax relation that
\citet{Juric2008} used for the blue stars was corrected for low-metallicities 
([Fe/H] $\sim$ $-$1.5\,dex), which is of minor difference from 
the photometric parallax we used for the halo stars.

The best-fit values for the disk fit are listed in Table~3.
We performed the disk fit using first jointly all three colour bins, then separately for each colour bin.
For the joint fit case, all the model parameters, excepted for 
the local density ($n_0$), are settled to be the same for stars in every bin,
as we expect a universal density profile for all stellar populations \citep{Juric2008}. 
All the parameters resulted from the joint fit appear to be very well constrained.
We explore the correlations between different parameter pairs in Fig.~\ref{cross2d}. 
The Figure shows the marginalized one- and 
two-dimensional PDFs of the model parameters. 
The correlations between different parameters are rather weak in general. 
We identify small correlations 
between several pairs of the parameters, such as ($H_1$, $H_2$), ($H_1$, $f_2$) 
and ($l_1$, $f_h$), and a strong degeneracy between the scale height and
the local normalised densities ratio of thick disk ($H_2$, $f_2$).

For the separate fits, we fit all parameters independently for each color bin. 
The individual fits can be served as a consistency check of our method.
From Table~3, we can declare that those best-fit solutions 
are generally consistent. The thin disk scale height varies around
$H_1 ~ \sim$ 320 pc, by $\pm 40$\,pc, which is 
consistent with the result from the  joint fit.
The variations of the disk scale lengths are relatively large, with 
$L_1~\sim$ 1.7 $-$ 2.8\,kpc and $L_2~\sim$ 2.9 -- 3.7\,kpc for the
thin disk and the thick disk, respectively. 
The large variations could be due to the limited ranges of Galactocentric distances 
on the Galactic plane for stars from both the XSTPS-GAC (for the limited depth) and 
the SDSS (for the poor sky coverage in low Galactic latitudes). 
Data with deeper depth and better sky coverage in the low Galactic latitudes
 (such as Pan-Starrs) may help to improve the situation.
The thick disk normalization $f_2$ and scale height $H_2$
appear also weakly constrained, with $f_2$ ranging from 7 to 14\,per\,cent and
$H_2$ from 730 to 1000\,pc. This is mainly due to the strong
correlation between these two parameters (see 
the  $f_2$ vs. $H_2$ panel in Fig.~\ref{cross2d}).  The halo normalisation $f_h$
is well constrained to $\sim$ 0.15\,per\,cent for the two G-type star bins. While
the value of $f_h$ is abnormally large for the late K-type star bin [1.5$ \le (g-i)_0 < $1.6]. 
This is mainly due to the fact that late K dwarfs in the halo are cut by our limiting magnitude ($r_0<$21\,mag).

We plot in Fig.~\ref{rnmodel} the star counts in the colour bin $0.5 \le (g-i)_0 < 0.6$\,mag
and magnitude bins, $r_0=$15 and 16\,mag of both the XSTPS-GAC 
and the SDSS data as a function of the Galactic latitude for example subfields 
with Galactic longitude 177\degr $ < l < $ 183\degr. 
The best-fit model is in good agreement with the observations,
with some small deviations.
In Fig.~\ref{resd}, we show the differences between the observed star counts, integrated from stars
in all the three colour bins and $r_0$ from 12 to 21\,mag, and 
the model predictions as a function of position on the sky. 
Different colours in the Figure indicate the values of the
ratio $(N_{\rm obs}-N_{\rm mod})/N_{\rm mod}$. For most 
of the fields, we do not see obvious deviation,
with residual smaller than 10\,per\,cent, i.e. 
$|(N_{\rm obs}-N_{\rm mod})/N_{\rm mod}| < 0.1$.

\subsection{Systematics}

The dispersions of the resultant parameters from both the jointly fit and individual fit 
are also listed in Table.~3, which could be used to denote the 
systematic errors of the corresponding parameters. 
The typical errors of the parameters are about 10\,per\,cent.  
Some of the parameters, i.e. the thin disk scale length $H_1$ and the thick disk 
local normalised densities ratio $f_2$, have relatively larger uncertainties, which are mainly due
to the limits of our data and the degeneracies between different parameters. 
Other dominant sources of the errors are (in order of decreasing importance):
1) the systematic distance determination uncertainties, 
2) the misidentification of binaries as single stars, 
3) the value of distance error, i.e., finite width of the photometric parallax relation,  
4) the contamination of non-dwarf stars, and 
5) the effects from disk warp and flare. Finally, the structure parameters for different 
stellar populations (colour bins) would be intrinsically different, which may also contribute to 
the dispersions.

Due to the absolute calibration errors of the photometric parallax relation, the distances of stars 
could be systematically over- or under- estimated. To check this effects, we
redo the fit by changing the distance scale by 􏰎15\,per\,cent, i.e., 
using the parallax relations 0.3\,mag brighter and 0.3\,mag fainter than Equation~(5).
The relative differences between 
the original resultant parameters and those derived after 
changing the distance scale by 15\,per\,cent are about 10 -- 15\,per\,cent.

Comparing to the single stars of the same colour, the binaries are 
brighter, i.e. of smaller absolute magnitudes. Thus if one model  
the Galaxy with no or less fraction of binaries, the resultant 
model would be more `compact' than it truly is. 
In the current work we select a binary fraction $f_b$=40\,per\,cent, 
which is an average binary fraction 
for field FGK stars \citep{Yuan2015b}. To check the possible effects 
of the binary fraction, we have tried two extreme 
values, the lowest value, $f_b$=0, assuming that all stars are 
single stars and the highest value, $f_b$=1, assuming that all stars
are binaries, to redo the fit. 
The relative differences between 
the original resultant parameters and those derived after 
changing the binary fraction to the extreme values are about 10\,per\,cent. 

When simulating the star counts for each colour bin, we assumed an error of distances
of 15\,per\,cent, which includes the effect of photometric 
uncertainties, dispersion of metallicities of disk stars and  
uncertainties of the \citet{Ivezic2008} photometric parallax 
relation. We redo the fit by changing the 
distance dispersion to 0 and 30\,per\,cent, respectively. 
The results show that changing the uncertainties of 
distance do not introduce any significant bias in the derived Galactic model parameters.     

The effects for the model parameters caused by the non-dwarf (i.e. subgiants and giants) contaminations
are similar as the binaries. From the Galaxy stellar population synthesis models, BESANCON
\citep{Robin2003} and TRILEGAL \citep{Girardi2005}, we find that the fraction of giants in the three 
chosen colour bins is no more than 1\,per\,cent and the fraction 
of sub-giants is less than 5\,per\,cent. Thus the systematics 
caused by the giant and sub-giant contaminations are likely to be negligible. 

In the current work we have ignored the effects of disk flare, as we assume that the stellar flare
is becoming significant at further distances \citep{Lopez2014}. While \citet{Derriere2001} and 
Amores et al. (2016, submitted) find that the flare starts at about 9 to 10\,kpc. Having a shorter
start of the flare could have an impact on the thin ansd thick disk scale lengths that we determine, 
which would make the values to be underestimated. 
   
\subsection{Comparisons with other work}

As stated in Section~4.1, our results of the halo model parameters are very similar
as those derived from \citet{Juric2008}, because of the similar method (stellar 
number density fits) and data (the SDSS) adopted in both work. 
The main differences in the halo fit between those from \citet{Juric2008}
and our work are that they adopt the $\chi^2$ fitting while we use the reduced
likelihood $L_r$; and they calculate the number densities 
for the entire sample in $(R,~Z)$ space while we do that separately for each line of sight. 
The result in our work confirms those from \citet{Juric2008}.

When constraining the disk model parameters,
our method and data are both quite different from those in \citet{Juric2008}
but the results are in agreement at a level of about 10\,per\,cent.
However, \citet{Juric2008} admit that their result suffers large uncertainties
from the uncertainty in calibration of the photometric parallax relation and the poor sky coverages 
for the low Galactic latitudes, which is not the case in our work.
Generally, our derived value of thin disk scale height, 322\,pc, is in the range of values, 
150--360\,pc, which resulted from the recent work by \citet{Bilir2008, Juric2008, 
Yaz2010, Chang2011, Polido2013, Jia2014} and \citet{Lopez2014}. 
Specially, this value is in good agreement with the canonical value of 325\,pc 
\citep{Gilmore1984, Yoshii1987, Reid1993, Larsen1996}. Our derived value of the
thin disk scale length, 2.3\,kpc, is 
consistent with those results found by \citet{Ojha1996, Robin2000, Chen2001, Siegel2002, Karaali2007, 
Juric2008, Robin2012, Polido2013, Lopez2014} and \citet{Yaz2015}, which ranges between 2 and 3\,kpc. 
We find a local thick disk normalization of 11 per\,cent.  In the different colour bins, 
this value varies between 7 and 14\,per\,cent, probably because of the degeneracy with the 
scale height. It is well in agreement with the values in the literature, which ranges between 7 and 13\,per\,cent \citep{Chen2001, Siegel2002, Cabrera2005, Juric2008, Chang2011, Jia2014}. 
The range of values for the thick disk scale height and scale length from the recent literature 
are respectively 600 -- 1000\,pc and 3 -- 5\,kpc \citep{Bilir2008, Juric2008, 
Yaz2010, Chang2011, Polido2013, Jia2014, Lopez2014, Robin2014}. The results deduced here, 
which have thick disk scale height of $\sim$800\,pc and scale length of 3.6\,kpc, are both in the
middle of the ranges of those values reported in the literature. Notice that there is a significant difference 
with the \citet{Robin2014} result for the thick disc. Their thick disk is modelled with 2 episodes, 
one of which has very similar parameters as the present result (their old thick disk). But their young thick 
disk is more compact, with smaller scale height and scale length. The difference can be due to the different 
shapes used (they use secant squared density laws and they include the flare) while in the present study 
the thick disk is a simple exponential vertically. 

\section{Substructures}

\begin{figure}
  \centering
  \includegraphics[width=0.48\textwidth]{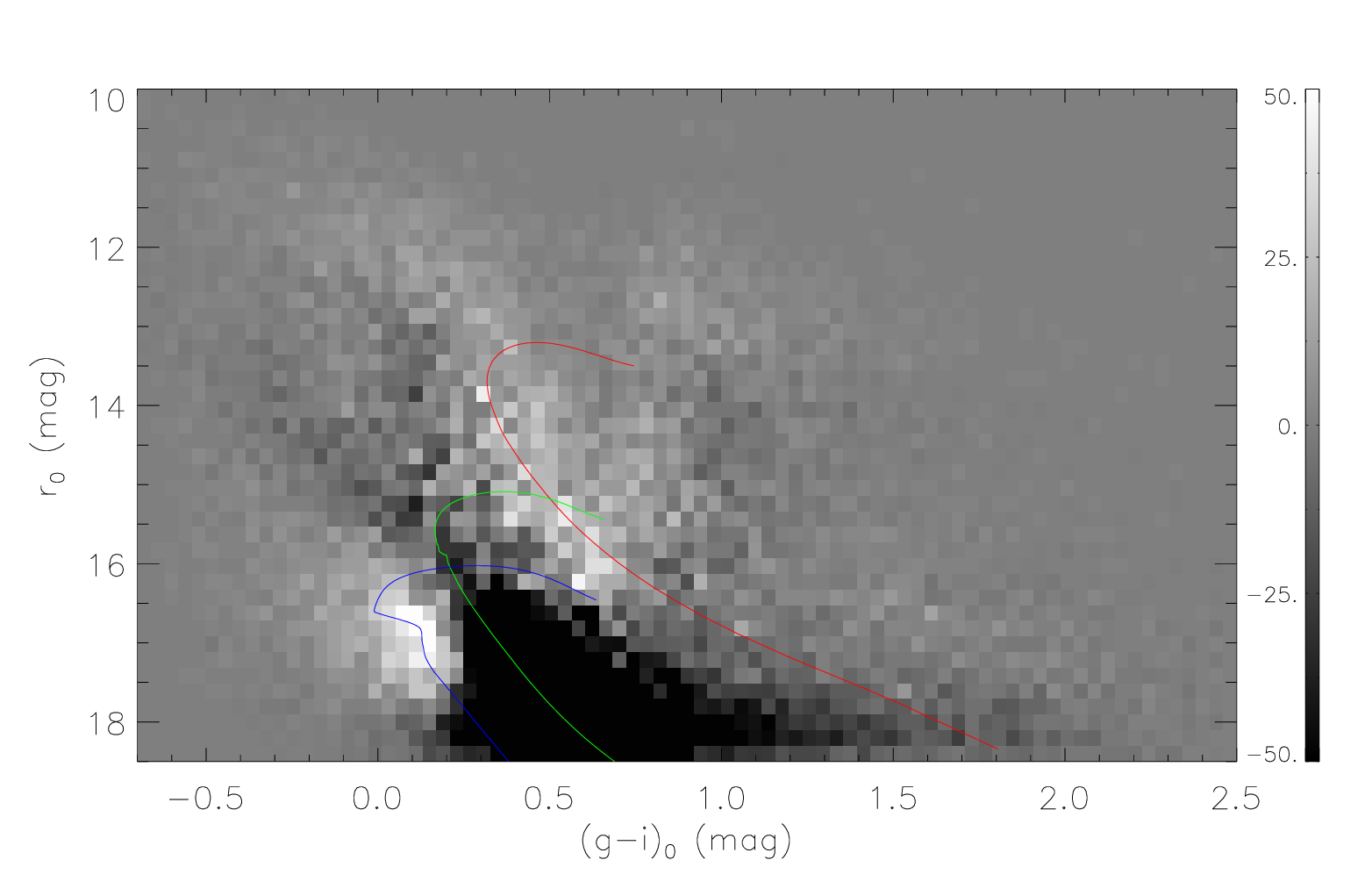}
  \includegraphics[width=0.48\textwidth]{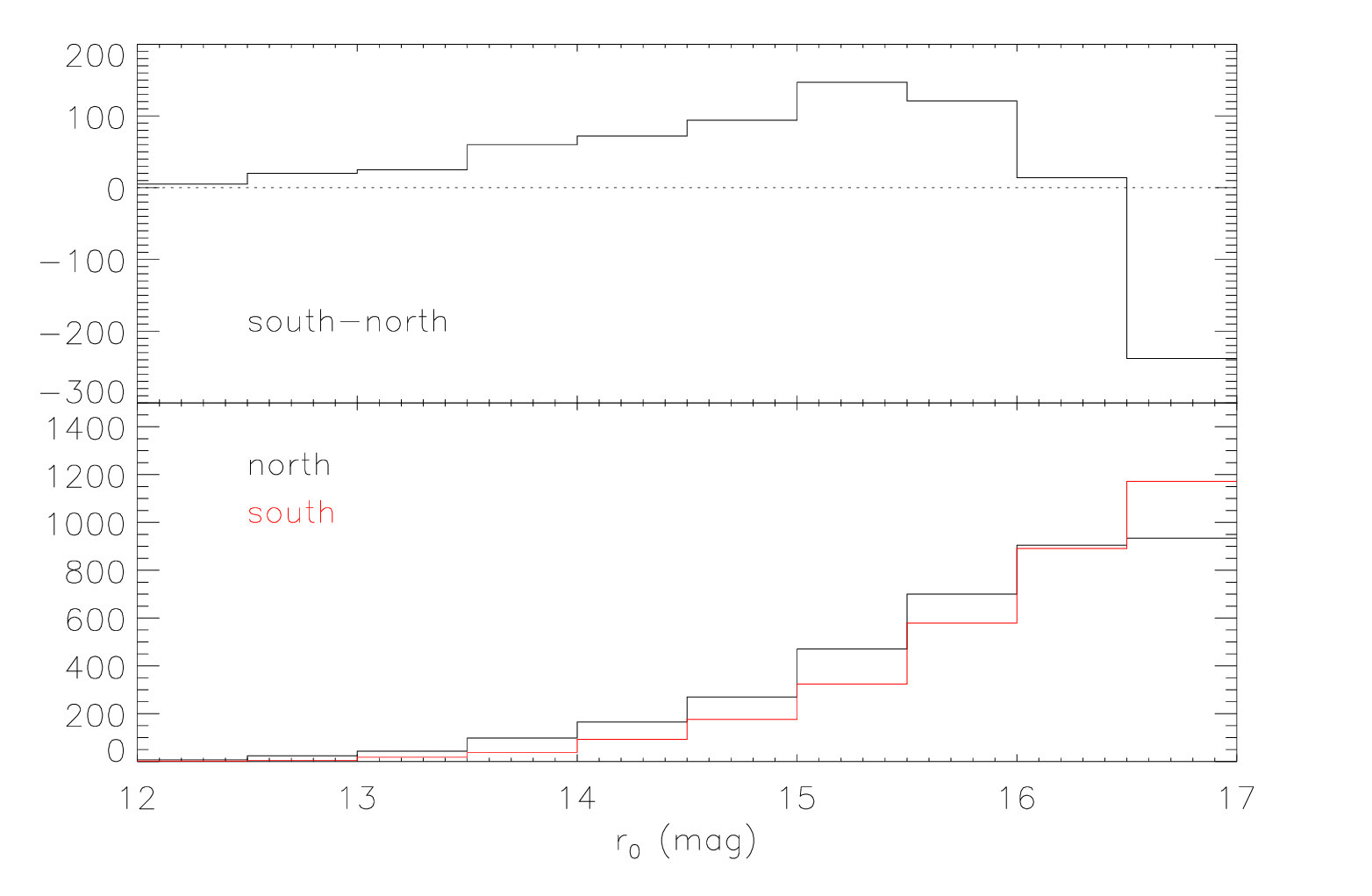}
  \caption{{\it Upper panel:} Differences of star counts in the $r_0$ vs. $(g-i)_0$ diagram for XSTPS-GAC data 
  in 2.5\degr $\times$ 2.5\degr patches of sky, i.e. ($l,b$) = (181.25\degr, $-$8.75\degr) minus ($l,b$) = (181.25\degr, 
  8.75\degr). The red curve which fits the white main-sequence pattern of the newly identified substructure
  is an isochrone with [Fe/H]=$-$0.2\,dex, age of 3.981\,Gyr and distance of 1.3\,kpc. The green curve that fits the black 
   main-sequence pattern of the ``north near structure'' is an isochrone with [Fe/H]=$-$0.5\,dex, age of 2.818\,Gyr and distance of 3.7\,kpc. The blue curve that fits the white main-sequence pattern of the ``south middle structure'' 
  is an isochrone with [Fe/H]=$-$0.5\,dex, age of 1.995\,Gyr and distance of 6.8\,kpc. 
{\it Bottom panels:} The $r$-band star counts in the colour bin $0.5< (g-i)_0<0.6$\,mag for the south patch
  ($l,b$) = (181.25\degr, $-$8.75\degr), north patch ($l,b$) = (181.25\degr, 8.75\degr) and 
  the subtraction of the two.}
  \label{l181}
\end{figure}

One of the main purposes of the Galactic smooth structure modelling  is to define the 
irregular structure of the Milky Way. Assuming that our model now 
reproduces the populations of the halo and the disk, 
we can investigate if any further structures are missing from our modelling. 
As seen in Fig.~\ref{resd}, most of the significant deviations are 
overdensities, i.e. $N_{obs} > N_{mod}$, except for 
a few subfields such as those located at ($l,b$)=(170\degr, 0\degr). The extinction in
these fields are large \citep{Chen2014}. It is very difficult to distinguish that whether it is a real  `hole'  
or it is caused by the selection effects or extinction correction errors. 
For the overdensities, we find three large scale structures,
which are located at different positions on the sky and appear at
different magnitudes. We describe them as follows.

The first large region where star counts are in excess is located at 
240\degr\ $<l<$ 330\degr~and  60\degr\ $<b<$ 90\degr.
It is a large and diffuse structure, which corresponds 
to the Virgo overdensity found by \citet{Juric2008}. 
For stars with $0.5 ~<~(g-i)_0~<~0.7$\,mag, the excess of star counts 
occurs at $r_0 \sim$20\,mag.
So the distance of the structure is between 9 -- 11\,kpc, which is 
consistent with the work of \citet{Juric2008}. A plausible explanation of the Virgo 
overdensity is that it is a result of a merger event involving the Milky Way and 
a smaller, low metallicity dwarf galaxy \citep{Juric2008}.   

The second significant structure is located at
170\degr\ $<l<$ 200\degr~and  10\degr\  $<b<$ 30\degr. 
The excess of star counts occurs at $r_0~\sim$ 16\,mag 
for stars with colour $0.5 ~<~(g-i)_0~<~0.7$\,mag, 
corresponding to a distance between 1.5--2\,kpc. 
This feature is consistent with the so-called 
``north near structure'' found by \citet{Xu2015}. 
\citet{Juric2008} also report an overdensity at
$R$ = 􏰉 9.5\,kpc and $Z$ =􏰉 0.8\,kpc, which
may be connected with this substructure.
\citet{Xu2015} report that this substructure 
represents one of the locations of peaks in the oscillations
 of the disk mid-plane, observed at 
about $\pm$15\degr Galactic latitude, toward the Galactic anticentre. 

The third star counts excess structure is located 
at 150\degr\ $<l<$ 210\degr~and  $-$15\degr\  $<b<$ $-5$\degr. 
We have not found any corresponding substructures to this 
feature from the literature.  
We need to check first that whether this
substructure is real or caused by the selection bias or 
high extinction effects. To test its existence in a more direct way, 
we examine the Hess diagrams constructed for a subfield located in
the area of the feature and for a control subfield which is outside the region.
We choose the control subfield with the same galactic longitude but opposite 
latitude as the selected subfield that includes the overdensity.
To avoid the effect of the selection bias of Sample\,C, 
we select data from the complete XSTPS-GAC photometric sample in the regions 
($l,b$) = (181.25\degr, $-$8.75\degr, south field) and
(181.25\degr, 8.75\degr, north field). Each region is of size 2.5\degr $\times$ 2.5\degr.
We correct the extinction of stars using the extinction map from \citet{Chen2015}
together with the extinction law from \citet{Yuan2013}.
The average extinction for the south and north fields
are respectively $A_r$ = 0.66 and 0.27\,mag. In Fig.~\ref{l181}, 
we show the differences between Hess diagrams 
and the star counts for stars with colour 0.5 $< (g-i)_0 < $ 0.6\,mag for the two regions. 
We see a white-black-white main-sequence pattern in the 
difference Hess diagram. The fainter white sequence which indicating that 
there are more stars on the south side
 and the middle black sequence indicating that 
there are more stars on the north side respectively correspond 
to the ``south middle structure'' and ``north near structure'' described  in \citet{Xu2015}.  
The bright white sequence corresponds to the newly found substructure. 
The $r$-band extinction in the these two fields are rather small. Even if we assume an error of 
20\,per\,cent in extinction, the uncertainties of colour $g-i$ and magnitude $r$ in the two 
fields are only $\delta _{g-i}$ = 0.09 and $\delta _r$ = 0.13 for the south field and 
$\delta _{g-i}$ = 0.04 and $\delta _r$ = 0.05 for the north field, respectively. 
The observed feature will not be erased by adjusting the reddening values. 
So we believe that the feature is real. The effect of selection function and inaccuracies 
in the reddening correction cannot cause the apparent overdensities.

The excess of star counts occurs at around $r_0~\sim 15$\,mag for stars in the 
colour ranges of 0.5 $< (g-i)_0 < $ 0.6\,mag, indicating a distance between 1 -- 1.5\,kpc. 
We select a series of isochrones from \citet{An2009}. The isochrone with [Fe/H] = $-$0.2\,dex 
and age of 3.981\,Gyr, shifted by a distance of 1.3\,kpc, fit well with the maximum overdensities of
the bright white sequence (see Fig.~\ref{l181} ). The young age and high 
metallicity are consistent with those of the field thin disk stars. Thus we believe that 
this feature is unlikely a substructure originated from the outer disk. The distance to the 
Galactic plane of this feature is about $Z~\sim~-$0.3\,kpc and that for the ``north near structure''
is about $Z~\sim~$0.5\,kpc. These two features, which show the significant
North-South asymmetry in the star number count distributions, are consistent with the 
vertical oscillations in the stellar density in the solar neighbourhood discovered
by \citet[see their Fig.~18]{Yanny2013}. 

\section{Summary}

Based on the data from the XSTPS-GAC and the SDSS, we have modelled the global 
smooth structure of the Milky Way. We adopt a three-component stellar distribution model.
It comprises two double exponential disks, the thin disk and the thick disk, and a
two-axial power-law ellipsoid halo. The stellar number density of halo stars in the colour bin
$0.5 < (g-i)_0 < 0.6$\,mag and the $r$-band differential
star counts in three colour bins, $0.5 < (g-i)_0 < 0.6$\,mag,
 $0.6 < (g-i)_0 < 0.7$\,mag and  $1.5 < (g-i)_0 < 1.6$\,mag, are
used to determine the Galactic model parameters. 
The best-fit values are listed in Table~2 and 3. In summary, the scale height and length
of the thin disk are $H_1$=322\,pc and $L_1$=2343\,pc,
and those of the thick disk are $H_2$=794\,pc 
and $L_2$=3638\,pc. The local stellar density ratio of thick-to-thin disk 
is $f_2$=11\,per\,cent, and that of 
halo-to-thin disk is $f_h$=0.16\,per\,cent.  
The axis ratio and power-law index of the halo are $\kappa=0.65$ 
and $p=2.79$. Our results are all well constrained and in good 
agreement with the previous works.


By subtracting the observations from our best-fit model, we find three
large overdensities. Two of them have been previously identified, 
including the Virgo overdensity in the Halo \citep{Juric2008}, which located 
at 240\degr\  $<l<$ 330\degr\ and  60\degr\  $<b<$ 90\degr\ with a distance between 9 -- 11\,kpc, 
and the so-called ``north near structure'' in the disk \citep{Xu2015}, which located 
at 170\degr\ $<l<$ 200\degr\ and 10\degr\ $<b<$ 30\degr\ with a distance between 1.5 -- 2\,kpc. 
The third structure, located at 150\degr\  $<l<$ 210\degr\ and   $-$15\degr\  $<b<$ $-5$\degr\ 
with a distance between 1 -- 1.5\,kpc, is a new identification. 
Through the Hess diagram examination, we conclude that 
it could not be a artifact caused  by extinction correction or selection effects. 
This feature, together with the ``north near structure'' confirms the earlier 
discovery of  \citet{Widrow2012} and \citet{Yanny2013} of a significant 
Galactic North-South asymmetry in the stellar number density distribution.

\section*{Acknowledgements}

We want to thank the referee, Prof. Gerry Gilmore, for his insightful comments.
This work is partially supported by National Key Basic Research Program of China
2014CB845700,  China Postdoctoral Science Foundation 2016M590014 and 
National Natural Science Foundation of China 11443006 and U1531244.
The LAMOST FELLOWSHIP is supported by Special Funding for Advanced Users,
budgeted and administrated by Center for Astronomical Mega-Science, Chinese
Academy of Sciences (CAMS). This research has made use of the Chinese 
Virtual Observatory (China-VO) resources and services.

This work has made use of data products from the Guoshoujing Telescope (the
Large Sky Area Multi-Object Fibre Spectroscopic Telescope, LAMOST). LAMOST
is a National Major Scientific Project built by the Chinese Academy of
Sciences. Funding for the project has been provided by the National
Development and Reform Commission. LAMOST is operated and managed by the
National Astronomical Observatories, Chinese Academy of Sciences.

Funding for SDSS-III has been provided by the Alfred P. Sloan Foundation, the
Participating Institutions, the National Science Foundation, and the U.S. Department
of Energy Office of Science. The SDSS-III web site is http://www.sdss3.org/.
SDSS-III is managed by the Astrophysical Research Consortium for the
Participating Institutions of the SDSS-III Collaboration including the University
of Arizona, the Brazilian Participation Group, Brookhaven National Laboratory,
Carnegie Mellon University, University of Florida, the French Participation Group,
the German Participation Group, Harvard University, the Instituto de Astrofisica de
Canarias, the Michigan State/Notre Dame/JINA Participation Group, Johns Hopkins
University, Lawrence Berkeley National Laboratory, Max Planck Institute for
Astrophysics, Max Planck Institute for Extraterrestrial Physics, New Mexico
State University, New York University, Ohio State University, Pennsylvania State
University, University of Portsmouth, Princeton University, the Spanish Participation
Group, University of Tokyo, University of Utah, Vanderbilt University, University of
Virginia, University of Washington, and Yale University.

\bibliographystyle{mn2e}
\bibliography{struct}
\label{lastpage}

\end{document}